%% file: paper.tex
\documentclass[conference]{IEEEtran}
\usepackage{cite}
\usepackage{amsmath,amssymb,amsfonts}
\usepackage{algorithmic}
\usepackage{graphicx}
\usepackage{textcomp}
\usepackage{xcolor}
\def\BibTeX{{\rm B\kern-.05em{\sc i\kern-.025em b}\kern-.08em
    T\kern-.1667em\lower.7ex\hbox{E}\kern-.125emX}}
\usepackage{listings}
\usepackage[linesnumbered,ruled,vlined]{algorithm2e}

\usepackage{breakurl}
\usepackage[breaklinks,hidelinks,bookmarks=false]{hyperref}
\hypersetup{draft}

\begin{document}

\title{Verifiable Smart Contract Portability
}

\author{\IEEEauthorblockN{Martin Westerkamp}
\IEEEauthorblockA{\textit{Service-centric Networking} \\
\textit{Telekom Innovation Laboratories, Technische Universit\"at Berlin}\\
Berlin, Germany \\
westerkamp@tu-berlin.de}
}


\maketitle

\begin{abstract}
With the advent of blockchain technologies, the idea of decentralized applications has gained traction.
Smart contracts permit the implementation of application logic to foster distributed systems that are capable of removing intermediaries.
Hereby, lock in effects originating from isolated data storage and central authorities are mitigated.
Yet, smart contracts deployed to a ledger generate dependencies on the underlying blockchain.
Over time, requirements regarding contract execution may detach from the utilized chain due to contradicting incentives and security or performance issues.
To avoid a novel form of lock in effect towards a host blockchain, we introduce a concept for smart contract portability that permits any user to migrate contract logic and state between blockchains in a flexible and verifiable manner.
As the Ethereum Virtual Machine (EVM) is supported by a multitude of blockchain implementations, it poses a common execution environment for smart contracts.
We provide a toolbox that facilitates smart contract portability between EVM-compatible blockchains without trust requirements in the entity executing the migration process.
To prove the concept's soundness, we transfer token contracts based on the ERC20 standard as well as applications containing dependencies to other smart contracts.
Our evaluation shows the validity of ported applications including their current states.
\end{abstract}

\begin{IEEEkeywords}
Blockchain, Distributed Ledger Technology (DLT), Smart Contract, Interoperability, Portability, Ethereum Virtual Machine (EVM)
\end{IEEEkeywords}

\input{helpers/solidity_highlighting}
\input{sections/introduction}
\input{sections/background}
\input{sections/concept}
\input{sections/evaluation}
\input{sections/related_work}
\input{sections/discussion}
\input{sections/conclusion}

\bibliographystyle{IEEEtran}
\bibliography{references}

\end{document}

%% file: helpers/solidity_highlighting.tex

\definecolor{verylightgray}{rgb}{.97,.97,.97}

\lstdefinelanguage{Solidity}{
	keywords=[1]{anonymous, assembly, assert, balance, break, call, callcode, case, catch, class, constant, continue, constructor, contract, debugger, default, delegatecall, delete, do, else, emit, event, experimental, export, external, false, finally, for, function, gas, if, implements, import, in, indexed, instanceof, interface, internal, is, length, library, log0, log1, log2, log3, log4, memory, modifier, new, payable, pragma, private, protected, public, pure, push, require, return, returns, revert, selfdestruct, send, solidity, storage, struct, suicide, super, switch, then, this, throw, transfer, true, try, typeof, using, value, view, while, with, addmod, ecrecover, keccak256, mulmod, ripemd160, sha256, sha3}, 
	keywordstyle=[1]\color{blue}\bfseries,
	keywords=[2]{address, bool, byte, bytes, bytes1, bytes2, bytes3, bytes4, bytes5, bytes6, bytes7, bytes8, bytes9, bytes10, bytes11, bytes12, bytes13, bytes14, bytes15, bytes16, bytes17, bytes18, bytes19, bytes20, bytes21, bytes22, bytes23, bytes24, bytes25, bytes26, bytes27, bytes28, bytes29, bytes30, bytes31, bytes32, enum, int, int8, int16, int24, int32, int40, int48, int56, int64, int72, int80, int88, int96, int104, int112, int120, int128, int136, int144, int152, int160, int168, int176, int184, int192, int200, int208, int216, int224, int232, int240, int248, int256, mapping, string, uint, uint8, uint16, uint24, uint32, uint40, uint48, uint56, uint64, uint72, uint80, uint88, uint96, uint104, uint112, uint120, uint128, uint136, uint144, uint152, uint160, uint168, uint176, uint184, uint192, uint200, uint208, uint216, uint224, uint232, uint240, uint248, uint256, var, void, ether, finney, szabo, wei, days, hours, minutes, seconds, weeks, years},	
	keywordstyle=[2]\color{teal}\bfseries,
	keywords=[3]{block, blockhash, coinbase, difficulty, gaslimit, number, timestamp, msg, data, gas, sender, sig, value, now, tx, gasprice, origin},	
	keywordstyle=[3]\color{violet}\bfseries,
	identifierstyle=\color{black},
	sensitive=false,
	comment=[l]{//},
	morecomment=[s]{/*}{*/},
	commentstyle=\color{gray}\ttfamily,
	stringstyle=\color{red}\ttfamily,
	morestring=[b]',
	morestring=[b]"
}

\lstset{
	language=Solidity,
	backgroundcolor=\color{verylightgray},
	extendedchars=true,
	basicstyle=\footnotesize\ttfamily,
	showstringspaces=false,
	showspaces=false,
	numbers=left,
	numberstyle=\footnotesize,
	numbersep=9pt,
	tabsize=2,
	breaklines=true,
	showtabs=false,
	captionpos=b
}

%% file: sections/introduction.tex
\section{Introduction}
Emerging blockchain technologies have leveraged a plethora of distributed applications that do not rely on central instances but are maintained in a decentralized fashion.
While the Bitcoin network, as proposed by Satoshi Nakamoto in 2008~\cite{nakamoto2008}, focuses on financial transactions independent from central authorities, general purpose blockchains such as Ethereum facilitate decentralization for a multitude of use cases~\cite{Buterin2014}.
To achieve this, smart contracts are implemented for expressing application-specific logic that is executed by all nodes participating in the network.
Hereby, third-party intermediaries are obsolete, as the shared contract logic is reproducible and transactions are cryptographically secure.

Applications utilizing smart contracts have evolved in various domains such as supply chain management~\cite{westerkamp2018, toyoda2017, korpela2017}, social networks~\cite{Chakravorty2017,biedermann2014}, identity management~\cite{sovrin2018, ali2016} and energy marketing~\cite{powerledger2018}. 
With a growing amount of smart contracts deployed to a single ledger, scalability becomes an increasing concern.
Furthermore, during a smart contract's life cycle, requirements may change.
For instance, the application scenario could shift from a public to a private environment or vice versa.
Similarly, the setting could require switching from a permissionless blockchain to a permissioned implementation.
Due to novel functionality or scalability, other blockchains may better suit a given use case.
Furthermore, blockchains hosting smart contracts may cease due to decreasing incentives, security issues or centralization tendencies~\cite{Beikverdi2015}, mitigating deployed applications' \textit{survivabiltiy}, as defined by Hardjono et. al~\cite{Hardjono2018}.
Several attack vectors have been proven to threaten blockchains~\cite{heilman2015,Tosh2017}.
In fact, incentives for maintaing a blockchain do not necessarily conform with those of smart contract users on the same chain~\cite{Shudo2018}.
This mismatch may lead to the requirement of porting a smart contract to another chain in order to mitigate involved security, availability or scalability issues.

Similarly, attack vectors affecting secure contract execution do not exclusively arise from underlying blockchains, but also originate from faulty contract implementations. The exposure multiplies when referenced libraries comprise vulnerabilities, as happened in Parity's multi-signature wallet implementation in 2017~\cite{Destefanis2018}. While referencing affected contracts in contract clones\footnote{https://github.com/ethereum/EIPs/blob/master/EIPS/eip-1167.md} could not recover locked cryptocurrency due to their native characteristic in Ethereum, it can restore asset contracts if users agree on a specific application replication.

Transferring smart contracts between blockchains requires a common execution environment.
While ledger implementations such as Neo\footnote{http://docs.neo.org/en-us/sc/introduction.html\#neovm} provide their own platform for smart contracts, the Ethereum Virtual Machine (EVM) has been adopted by a multitude of blockchains outside the Ethereum universe such as Hyperledger Sawtooth\footnote{https://www.hyperledger.org/projects/sawtooth}, Hyperledger Fabric\footnote{https://github.com/hyperledger/fabric-chaincode-evm}, Hyperledger Burrow\footnote{https://www.hyperledger.org/projects/hyperledger-burrow}, Quorum\footnote{https://www.jpmorgan.com/global/Quorum}, Qtum\footnote{https://qtum.org/en} and Counterparty\footnote{https://counterparty.io/docs/faq-smartcontracts/}.
Thereby, a smart contract written in a high level programming language like Solidity and compiled to EVM-compatible bytecode is deployable to any blockchain supporting the EVM.

In this paper, we propose a concept for verifiable smart contract portability targeting EVM-compatible blockchains.
We present a tool that implements all required steps of the migration process, namely execution code retrieval, state reconstruction and generation of deploy-ready bytecode.
Furthermore, any blockchain participant is enabled to verify the migration's validity by comparing the roots of the original and novel state Merkle trees.
Therefore, no trust in the executing party is required as both smart contract logic and state are publicly verifiable.
All transformations are performed in bytecode.
As a result, the contract's high level source code is not required for portability operations, enabling smart contract forks that are independent from contract owners or primary deploying instances.


%% file: sections/background.tex
\section{Background}
\subsection{Blockchains, Smart Contracts and the EVM}
Blockchains provide an immutable peer-to-peer storage of digitally signed transactions that permit, for instance, exchanging assets without requiring any intermediaries.
In addition to simple transactions sending assets from one party to another, Bitcoin supplies a scripting language for implementing additional functionality such as multi-signature transactions~\cite{Tschorsch2016}.
However, the scope of Bitcoin scripts is limited to very basic procedures \cite{Bartoletti2017}.

\begin{figure} 
	\centering
	\includegraphics[width=.49\textwidth]{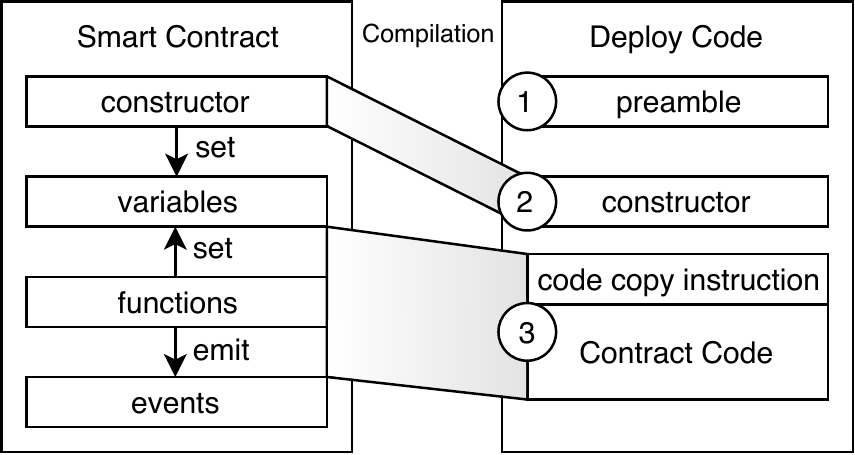}
	\caption{High level Solidity code is compiled to deployment code which is embedded into a transaction}
	\label{fig:compilation}
\end{figure}

To overcome this limitation, Ethereum was proposed as an alternative blockchain implementation that provides a quasi Turing-complete virtual machine, the EVM\cite{Buterin2014}.
Programs written for the EVM, called smart contracts, are typically developed using a high-level programming language such as Solidity\footnote{https://github.com/ethereum/solidity} or Vyper\footnote{https://github.com/ethereum/vyper}.
These high-level contracts are then compiled to bytecode that is interpreted by the EVM.
The resulting program code consists of multiple components, as exemplified in Figure~\ref{fig:compilation}.
First, deployment-specific code is generated that instructs the EVM to allocate memory, store contract code and execute the constructor~\cite{Wood2014}.
Second, the constructor sets the initial state.
Because it is only executed during the contract's deployment, its bytecode is not included in the application logic, namely the contract code.
Nevertheless, it changes the contract instance's state according to the declared instructions.
Third, functions implemented in Solidity permit executing calculations, accessing the application's storage and triggering other internal or external procedures on the same chain.
Accordingly, functions are compiled to runtime code by the Solidity compiler which is a subset of the deploy code that is executed when deploying smart contracts to the blockchain.
While the EVM does not natively support functions, referenced code can be called using the original function's signature which is defined as the first four bytes of the method's name and parameter hash~\cite{Atzei2017}.
Within the deploy code, the EVM is instructed to copy the compiled runtime code, it will therefore be part of an account in the blockchain.
The compiler also calculates storage locations for each variable set.
As the EVM uses key value mappings for allocating variables in the storage tree, single variables are numbered sequentially by the Solidity compiler.
Values stored in mappings and arrays are allocated according to their keys' \textit{keccak256} hash.
In addition to constructor and contract code, a preamble is generated during the compilation process that allocates memory and calls required deployment instructions.

Every operation in the EVM requires a specific amount of gas for its execution \cite{Wood2014}.
Blockchains implementing the EVM specify a gas limit for each block to determine an upper bound of execution steps.
In Ethereum, transaction senders must include cryptocurrency in each transaction to bid for gas that is needed for its execution.
Hereby, a market is created for the limited gas supply, leading to increasing gas prices with a growing number of transactions.

Unlike Bitcoin, which relies on unspent transaction outputs (UTXO) to facilitate asset transfers~\cite{nakamoto2008,Tschorsch2016}, Ethereum follows a state-based account structure.
Two forms of accounts exist: user accounts, also referred to as externally owned accounts (EOA), and smart contracts.
Each account maintains its own state that is stored in a Merkle Patricia Tree (trie) for fast and secure verifiability (cf. Section~\ref{subsec:merkle}).
During contract execution, not only the internal state may change, but also external smart contracts deployed on the same chain can be triggered.
In addition, events can be emitted to trigger off-chain processes such as user interface updates, or to provide a different kind of storage that cannot be accessed from smart contracts but save gas costs in contrast to state storage operations~\cite{westerkamp2018}.
Events are not stored by contract but are part of the receipt Merkle tree that is attached to each block in Ethereum.
Therefore, to retrieve past events, nodes need to iterate over all blocks from the most current block to the block that holds the contract's deployment transaction.
To increase efficiency, Ethereum utilizes bloom filters to check if a block holds relevant events~\cite{Wood2014}.

\subsection{Merkle Trees}
\label{subsec:merkle}
Bitcoin and Ethereum utilize Merkle trees for efficiently verifying transactions and states respectively.
Merkle trees permit proving the inclusion of data in a tree of hashes without requiring knowledge about the underlying data set~\cite{Merkle1980}.
In fact, such proofs can be performed on pruned trees where only the path from root to the target hash and child nodes at the first level are included.

Ethereum relies on modified Merkle Patricia Trees which permit efficiently persisting data in a map structure with arbitrary length~\cite{Buterin2014}.
Data in the tree is encoded using the recursive length prefix encoding (RLP)~\cite{Wood2014}.
For efficient traversal, three types of nodes are introduced: leaf, extension and branch.
As illustrated in Figure~\ref{fig:trie}, each block in Ethereum stores three distinct Merkle trees to save transactions, receipts and state~\cite{merkling-in-ethereum}.
While branches comprise up to 16 references, Figure~\ref{fig:trie} only depicts a subset for clarity reasons.
The state trie holds all accounts and contract data, with a dedicated Merkle tree linked to each account, persisting maintained variables.
In the state trie, an account is identified by its address's hash that maps to an array of nonce, balance, storage root and code hash, where the nonce indicates the account's transaction count.

%% file: sections/concept.tex
\section{Portable Contracts}
Migrating smart contracts between blockchains poses multiple requirements.
First, both source and target chain need to be able to interpret the given contract code.
We have identified a multitude of distributed ledgers implementing the EVM which provides a common execution environment for such applications.
Second, the contract state in a given block of the source chain should be maintained and transferred to the target chain.
Third, the correctness of the ported application should be verifiable by third parties.

\begin{figure*} [h]
	\centering
	\includegraphics[width=.85\textwidth]{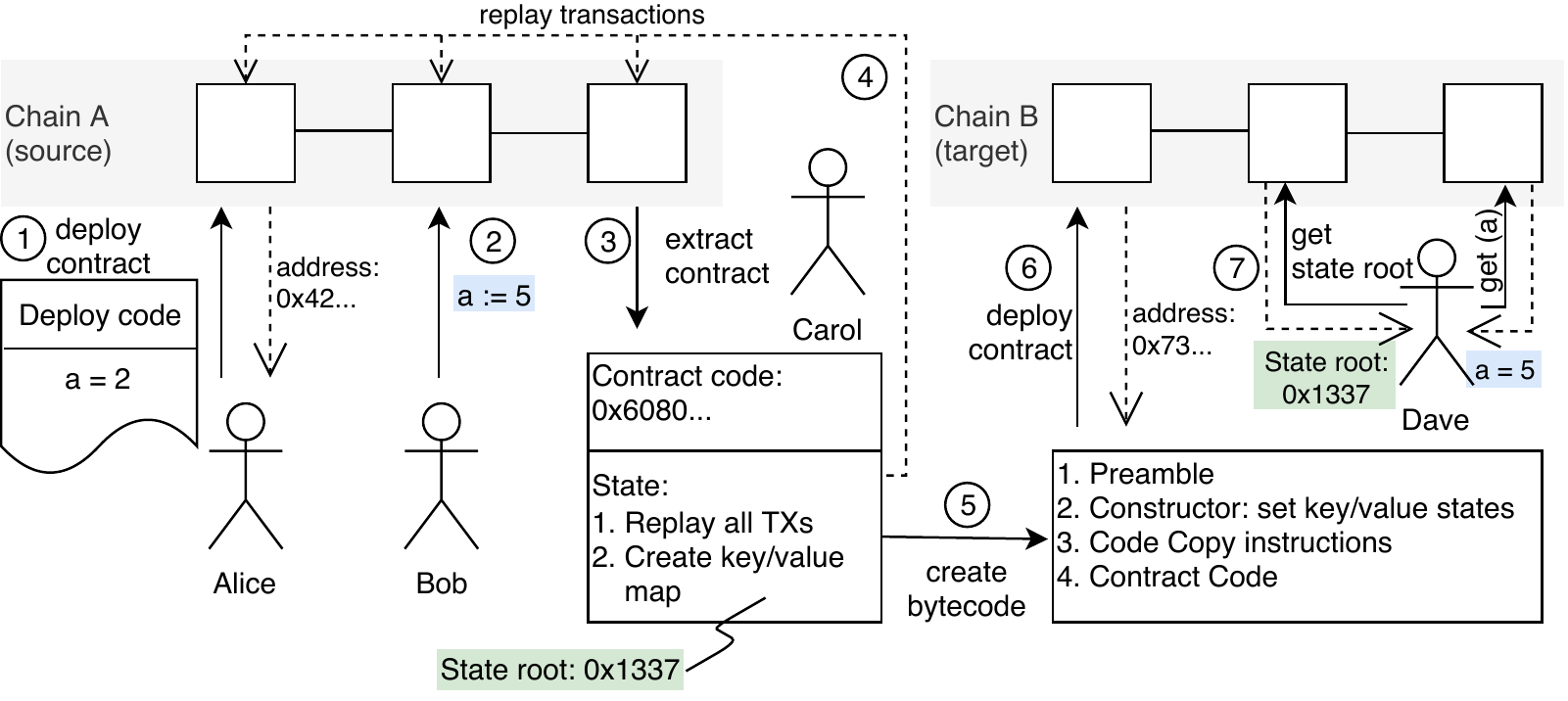}
	\caption{A smart contract is deployed and used on blockchain A.  It's state is reconstructed to create respective migration code for blockchain B. The ported contract is verified by a third party that has not been involved in the migration process.}
	\label{fig:general_concept}
\end{figure*}

To facilitate smart contract forks and portability to blockchains that are independent from the source chain, we propose a migration process requiring multiple steps, as illustrated in Figure~\ref{fig:general_concept}.
Therein, several entities interact with the smart contract.
Namely, Alice develops the contract in a high level language such as Solidity, compiles it to bytecode and deploys it to blockchain A.
Thereafter, Bob executes a setter function and modifies a state variable.
In order to port the contract to another blockchain, Carol retrieves the runtime code and state.
These components are used for constructing deploycode that stores identical contract code on the target blockchain and sets all state variables equivalent to the source contract's current state.
Finally, while not being involved in the migration process, Dave is enabled to verify the equality of both execution code and state between source and target chain by comparing their state trie hashes.

\begin{figure} [h]
	\centering
	\includegraphics[width=.49\textwidth]{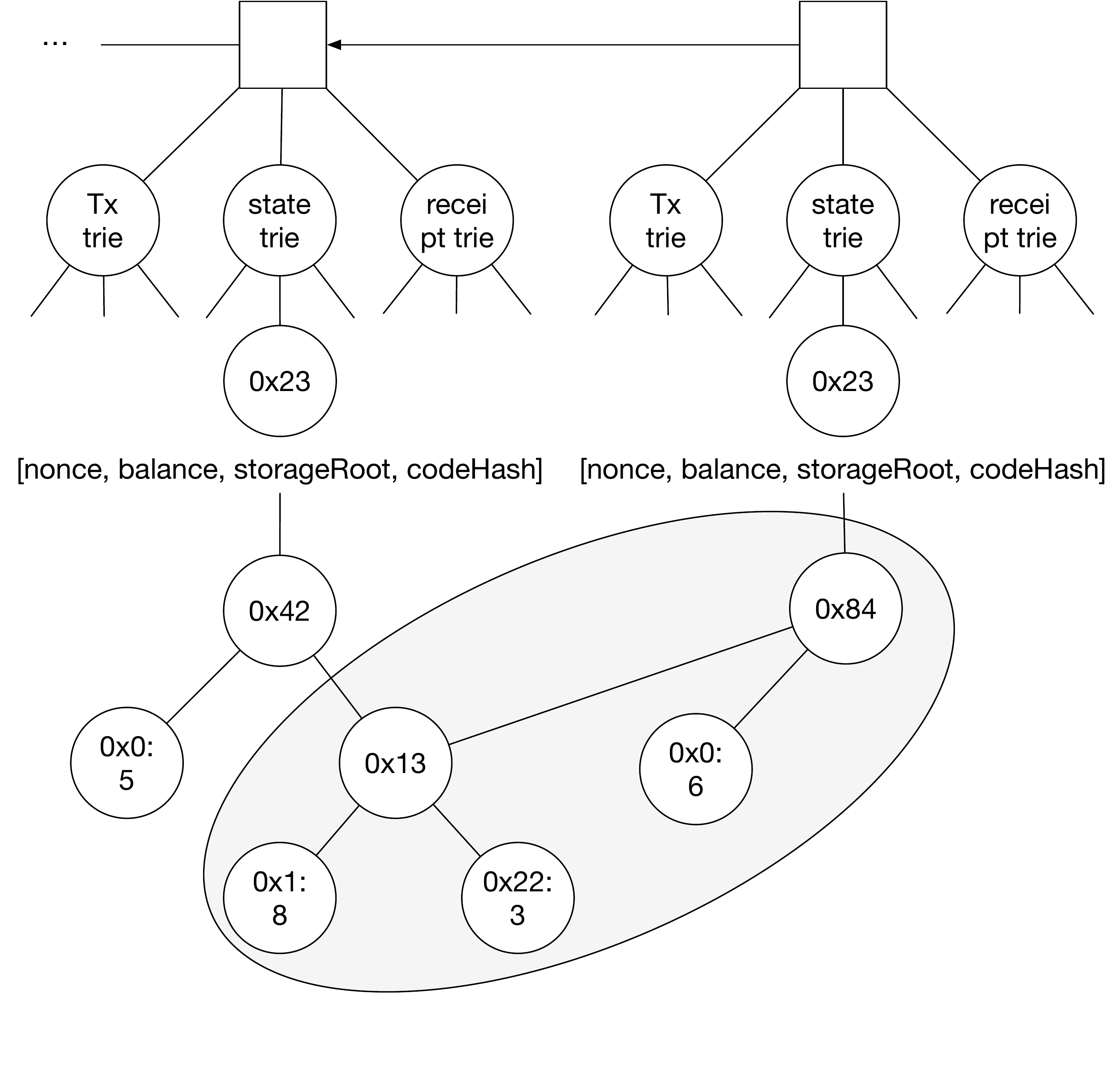}
	\caption{Each block in Ethereum holds three distinct Merkle trees for transactions, states and receipts. Accounts may maintain discrete storage tries.}
	\label{fig:trie}
\end{figure}

Migrating smart contracts from one blockchain to another should not be restricted to contract owners or developers, as they depict central entities, mitigating the original idea of distribution and independence of central authorities. 
To achieve this target, all operations are conducted in bytecode, providing flexibility and security, as the contract code is available to every participant in the network and the transferred code will be verifiably identical to the source contract.
The deployed contract code can be retrieved directly from Ethereum client nodes such as geth or Parity.
A migrated contract constitutes an application fork since the original contract remains unaffected.
However, in case the migrating entity controls the source contract, it may be destructed by its owner if intended.

To reconstruct the contract's state, all contained variables have to be observed from the ledger.
Each smart contract maintains a unique state tree holding all state variables.
Figure~\ref{fig:trie} presents an example of the impact on the state tree when modifying a single variable in a smart contract.
Changing the value associated with key 0x0 only affects its respective node and referencing nodes while the remaining tree is unaffected.
The current state that is required for the contract's migration is highlighted in grey in Figure~\ref{fig:trie}.
In order to provide an example of different kinds of variables available in Solidity and to showcase how they are persisted in the state trie, we implement a simple smart contract that is depicted in Listing~\ref{lst:simple_contract}.
While retrieving the variables directly from the Merkle tree by accessing the node's underlying key/value database would be sufficient for single variables (cf. Listing~\ref{lst:simple_contract}, Line~\ref{line:single}), as keys are numbered sequentially by the Solidity compiler\footnote{https://solidity.readthedocs.io/en/develop/miscellaneous.html\#layout-of-state-variables-in-storage}, this is not the case for mappings and arrays (cf. Listing~\ref{lst:simple_contract}, Line~\ref{line:mapping}) or bytecode that was created using different compilers.
To prevent Denial-of-service (DoS) attacks that create Merkle trees with maximum branch length, keys are hashed by the EVM so that the resulting key for allocating the variable  in the Merkle tree cannot be easily manipulated~\cite{ethereumdesign2018}.
Therefore, when retrieving the keys from the Merkle tree directly to create storage instructions, they would be hashed again during the deployment, destroying the original mapping.
By this means, the information stored in the database is not sufficient for restoring its state on a another blockchain.

\pagebreak
Ethereum node implementations such as geth\footnote{https://github.com/ethereum/go-ethereum/wiki/Management-APIs\#debug\_tracetransaction} and Parity\footnote{https://wiki.parity.io/JSONRPC-trace-module} provide a debug mode that permits replaying past transactions.
By this means, transactions are executed locally to record state changes.
State differentials do not only include affected values but also their respective storage location in the state trie.
A prerequisite for debugging is a full node that does not perform trie pruning, as the entire transaction history has to be available.
To trace the outcome of transactions, the blockchain state that existed before the transaction was mined is reconstructed and the transaction is executed, providing the corresponding outcome.
As a result, all state changes can be observed, providing information about the keys used in mappings as well as single variables' keys used to allocate it in the contract's key/value store.
We utilize this mechanism to reconstruct the smart contract's current state.
First, all transactions sent to the smart contracts have to be discovered, including its initial deployment transaction.
To deploy contracts, the contract code is included in a transaction that instructs the EVM to copy the respective code to a new account.
Information about deployed contracts is provided in the transaction's receipt.
Transaction receipts are stored in a unique Merkle tree referenced in each block that has been mined.
As receipts are stored independently from the referenced account, all blockchain transaction receipts have to be explored from the current block back to the block that holds the deploying transaction in order to examine if they are related to the target contract.
Due to the rapidly growing ledger size of the Ethereum main chain, iterating over all transactions can be computationally intensive, depending on the contract's time of deployment.
To permit accessing all relevant transactions, we extract transaction data into an external, structured data storage.
Instead of querying a blockchain node directly, the structured data is utilized for fast information retrieval.
While maintaining an external transaction database leads to significant performance improvements, it is not mandatory for successfully migrating smart contracts.
Thereafter, all relevant transactions are fetched from this database and replayed sequentially using an Ethereum archive node to reconstruct the final state, as illustrated in Figure~\ref{fig:general_concept}.
State changes resulting from transactions are stored in a key/value store with the original key inplace instead of the hashed key that is persisted in the Merkle tree.
As the transactions are replayed sequentially, we eventually obtain the final contract variable state.

\begin{lstlisting}[language=Solidity, caption=Example contract containing three different kinds of variables, xleftmargin=5.0ex,label=lst:simple_contract,escapechar=|]
import "./ReferencedContract.sol";

contract SimpleContract {
	uint256 a;|\label{line:single}|
	mapping (uint256 => uint256) b;|\label{line:mapping}|
	ReferencedContract referncedContr;|\label{line:reference_variable}|
	
	constructor(address _referncedContr) public {
		a = 42;
		b[13] = 21;
		referncedContr = ReferencedContract(_referncedContr);
	}
	
	function setA(uint256 _a) public {
		a = _a;
	}
	
	function setB(uint256 key, uint256 val) public {
		b[key] = val;
	}
	
	function setRefVar(uint256 _var) public {
		referncedContr.setVar(_var);
	}
}
\end{lstlisting}

\subsection{Deployment strategies}
After receiving the contract's code and state, the deployment instructions for applying the smart contract logic and state to another chain is generated in bytecode.
Depending on the contract code and state size, we propose two distinct deployment strategies.
As a smart contract's state tends to grow over time due to an increasing amount of variables stored in maps and arrays, storing the entire state during deployment would exceed Ethereum's gas limit.
Therefore, the deploying transaction has to be aligned with the gas limit within the target chain.

\subsubsection{Single Contract}
In case the gas used for deploying the contract code and executing all instructions required to recover its state from the source chain are smaller than the target chain's gas limit, the migration is conducted within a single transaction.
For that purpose, we utilize the fact that the constructor is, while being part of the deploy code, excluded from the actual contract code that is persisted on the blockchain, as illustrated in Figure~\ref{fig:compilation}.
Therefore, we generate a new constructor that recreates the source contract's state from our produced key/value map.
To create respective deploy code, first, a static preamble is applied to set the correct memory pointer, as demonstrated in Algorithm~\ref{pseudocode_deploy_code}, Line~\ref{pseudocode_preamble}.
Thereafter, the storage instructions for each key/value pair observed from the state are generated and appended to the bytecode (cf. Algorithm~\ref{pseudocode_deploy_code}, Lines~\ref{pseudocode_loop_begin}-\ref{pseudocode_loop_end}).
Here, both key and value are pushed to the EVM's stack before calling the \texttt{SSTORE} opcode.
Furthermore, the byte size has to be calculated in order to select correct \texttt{PUSH} opcodes which range from 1 to 32 bytes in the EVM~\cite{Wood2014}.
To observe the respective opcode, we utilize the fact that all \texttt{PUSH} instructions are represented sequentially in bytecode, so that the length can be added to its lowest operation, as presented in Algorithm~\ref{pseudocode_deploy_code}, Lines~\ref{calculate_byte_length} and~\ref{add_byte_length}.
After setting all variables, the contract code copy instructions have to be generated.
As the location of the contract code within the deployment code is required, it has to be calculated with regard to the constructor and contract code size, before adding the corresponding bytecode.
In the end, the contract code is included as it was retrieved from the source chain.

Sending a transaction including the generated deploy code to the target chain results in the creation of a new smart contract that holds the identical contract code and state.
The equality of smart contracts can be verified by any user and is not limited to the transferring entity.
For verification, first, both source and target contract are retrieved from the respective blockchain nodes and compared.
Second, both contracts' Merkle trees are retrieved from the nodes' databases.
In fact, only the trees' root hashes need to be compared to guarantee the equality of both states.
As a result, smart contracts can be forked either on the same blockchain or between EVM-compatible chains.
Due to the facilitated verifiability, no trust in the entity performing the transferring process is required. 
\begin{algorithm}[h]
	\DontPrintSemicolon
	\SetKwInOut{Input}{input}\SetKwInOut{Output}{output}
	\Input{%
		$S -$ Set of state variable key/value pair\newline
		$C -$ Contract Code
	}
	\Output{$D -$ Deploy Code\newline}
	
	\Begin{
		\tcc{Add static preamble}
		$D\leftarrow 0x608060405234801561001057600080fd5b50$\;\label{pseudocode_preamble}
		\ForEach{$s\in S$}{\label{pseudocode_loop_begin}%
			$l_{key}\leftarrow byteLength(s_{key})$\;
			$l_{value}\leftarrow byteLength(s_{value})$\;\label{calculate_byte_length}
			$D\leftarrow D~.~opcode(PUSH1)+l_{value}-1$\;\label{add_byte_length}
			$D\leftarrow D~.~opcode(s_{value})$\;
			$D\leftarrow D~.~opcode(PUSH1)+l_{key}-1$\;
			$D\leftarrow D~.~opcode(s_{key})$\;
			$D\leftarrow D~.~opcode(SSTORE)$\;
		}\label{pseudocode_loop_end}
		$l_{code}\leftarrow byteLength(C)$\;
		$D\leftarrow D~.~opcode(PUSH1) + l_{code}$\;
		$D\leftarrow D~.~l_{code}$\;
		$D\leftarrow D~.~DUP1$\;
		$D\leftarrow D~.~byteLength(D)$\;
		$D\leftarrow D~.~CODECOPY$\;
	}
	\caption{Create Deploy Code}\label{pseudocode_deploy_code}
\end{algorithm}\DecMargin{1em}

\subsubsection{Proxy Contract}
In case the source contract or its state are too large for deployment within a single transaction, the reconstruction process has to be split into multiple transactions.
While adding a function to the original contract code would permit iteratively setting the correct state using transactions distributed over numerous blocks to avoid exceeding the gas limit, such alteration mitigates the ported contract's verifiability.
Therefore, three contracts are generated that perform distinct tasks:
\begin{enumerate}
	\item The \textit{logic contract} contains the source contract code without any alterations.
	\item The \textit{proxy contract} holds all variables and redirects all method calls to the logic contract.
	\item The \textit{initialization contract} is authorized by the proxy contract to set its state variables and destroyed when the transfer operation is completed.
\end{enumerate}
We utilize the EVM's capability of calling a remote contract's function while operating on the calling contract's state, by using the \texttt{delegatecall} opcode~\cite{Wood2014}.
Hereby, the deployed contract code is only responsible for maintaining the source contract's logic but is independent from its state.
Instead, all state variables are managed within a proxy contract that implements two functions: setting its state for porting the source state and a fallback function that is executed every time an unknown method is called.
The first function is used to facilitate dividing the state recreation into multiple transactions.
As it is crucial from a security perspective that manipulating the contract state is only permitted during an initialization phase, we define an authorized contract responsible for this task.
Only transactions sent from this initialization contract are accepted.
After all variables have been set in the proxy contract, the initialization contract is destroyed, so that it cannot be used for future variable modifications.
Through this mechanism, only the original contract state is part of the proxy contract, ensuring verifiability by maintaining an equal Merkle tree compared to the source state.
Therefore, the initialization contract's address has to be hard coded into the contract code in order to avoid alterations to the state.
The proxy contract's fallback function is called every time the calling method's signature is not recognized by the EVM.
It forwards the method call to the logic contract including its signature and parameters.
While the logic contract executes the initiated function, it operates on the proxy contract's state.

\begin{figure} 
	\centering
	\includegraphics[width=.49\textwidth]{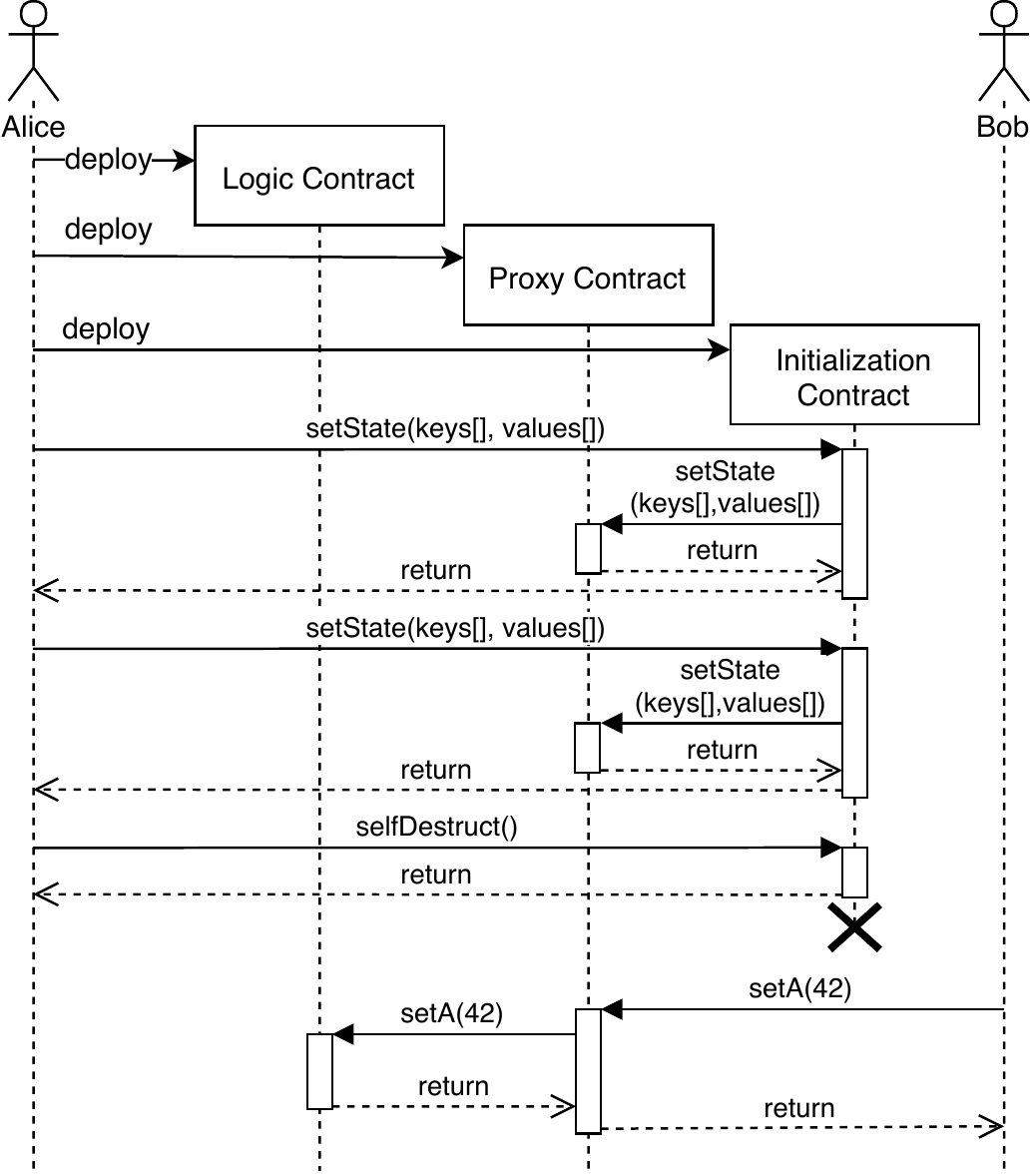}
	\caption{Example of deployment and interaction process with a proxy contract}
	\label{fig:proxy_contract}
\end{figure}

The three distinct contracts should be deployed in the same order as depicted in Figure~\ref{fig:proxy_contract} in order to embed correct reference addresses.
Bidirectional dependencies of proxy and initialization contract either require the initialization contract to maintain a respective state variable or the target address to be precalculated.
In the first case,  a function in the initialization contract permits setting the reference contract.
Instead, to decrease gas cost resulting from additional deploy code and transactions, the referenced contract's address can be calculated by it the sender's address and transaction nonce (transaction count) \cite{Wood2014}.
The calculated address is set in the contract's bytecode, preventing the requirement for any further setup.

After the contract migration is completed, users should address the proxy contract which forwards the method call to the logic contract, but provides the state that was ported previously.
The source contract remains unaffected from the smart contract fork.
%

\subsection{Dynamic Contract References}
Smart contracts are not limited to isolated state variables, but may refer to other contracts deployed to the same chain, as exemplified in Listing \ref{lst:simple_contract}, Line \ref{line:reference_variable}.
These references are mutable and can be adjusted on demand. Thus, the correct execution of smart contracts also relies on associated contracts.
To maintain such connections during the migration process, first, references are extracted by iterating over all state variables and checking them for valid addresses.
Thereafter, it is examined if the addresses hold any contract code. If nothing is retrieved, we assume the address refers to an externally owned account. Otherwise, it depicts a smart contract that is migrated recursively.
By this means, not only the target contract is migrated, but also all dependencies.
Such referenced smart contracts can be represented in a tree-like structure that depicts all required migrations.
As redeployed contracts retrieve new addresses, respective references need to be updated accordingly in the target contract.

Following historic migrations of referenced contracts, some dependencies may have already been migrated to the target chain.
Depending on the use-case, it may be beneficial to either redeploy the contract, for instance if the state has diverted due to its utilization during a given time disparity, or maintain the already deployed instance in case it is intended to rely on its current state.
To validate a migration process, it is not sufficient to only approve the root contract, but also all related contracts need to be considered.
For this purpose, transparency indicating if all related contracts have been redeployed during migration or if it relies on other contract instances is required.

Replacing contract references in the state also modifies the resulting state trie.
Therefore, the correct state transfer cannot be validated by comparing the state trie root hashes of source and target contracts in this case.
As a result, all values need to be verified according to the claimed referenced contracts and unaffected state variables.
\subsection{Static Contract References}
Despite dynamic references to smart contracts, addresses may also be encoded in the contract code itself.
Respective addresses can either be declared directly in high-level code such as Solidity or woven into the bytecode at compile time.
For instance, when utilizing a library, such contracts are mostly constant, already deployed to the blockchain and do not change over time.
In order to create a link between the newly deployed contract and the library instance, smart contract tools like truffle\footnote{https://github.com/trufflesuite/truffle} permit specifying the library's address during the deployment process. It is then inserted into the bytecode accordingly by the Solidity compiler.
Similarly to dynamic references, static links need to be migrated recursively to the target chain in order to maintain existing contract connections.
To observe present static references, the bytecode is analyzed for dependencies using the contract analysis toolbox Mythril\footnote{https://github.com/ConsenSys/mythril-classic}.
After migrating all corresponding contracts, their source addresses are replaced with the newly retrieved target addresses.

As substituting addresses in the contract code leads to a failure in validation when comparing source and target code, a map of replaced addresses needs to be supplied.
Using this map, first, the correct deployment of contract code can be validated.
Second, all migrated contracts are checked recursively based on the address map.
Hereby, the target contract is migrated including all relations in a transparent and verifiable manner.
	


%% file: sections/evaluation.tex
\section{Evaluation}
For evaluating the feasibility of the presented smart contract portability concept, we provide a toolbox that provides all proposed functionality.
It contains multiple components targeting each step during the migration process and is published under an open source license on GitHub\footnote{https://github.com/informartin/VeriSmart}.

All components that retrieve contract code  or related transactions and execute transactions utilize Ethereum nodes by using the web3\footnote{https://github.com/ethereum/web3.js/} library or direct JSON RPC queries.
Methods that target debug operations such as replaying transactions are implemented specifically to a given client.
Thus, they are not included in web3, but have to be called explicitly.
Currently, Parity and geth as well as their forks such as Quorum are supported.
While the state is analyzed for observing possible contract references within our implementation, we rely on the security analysis tool Mythril for extracting static dependencies.
As the storage root of smart contracts cannot be retrieved from the given node implementations immediately, we supply a validation tool for underlying database implementations.
The key/value database LevelDB is accessed directly, as geth utilizes it for storing chain data.

\subsection{Token Contracts}
As a major use-case for applications on Ethereum are token contracts \cite{victor2019measuring}, we randomly choose a smart contract\footnote{Contract address: 0x2A05d22DB079BC40C2f77a1d1fF703a56E631cc1} that implements the ERC20 token standard\footnote{https://eips.ethereum.org/EIPS/eip-20} to prove the presented concept's validity.
A prerequisite for transaction tracing is an archive node that does not perform any state trie pruning.
Therefore, we synchronize the Ethereum mainchain using the Parity client.
Retrieved contracts should then be ported to a public testnet and a private, permissioned chain using several node implementations.

For the selected contract, we observe 3,179 transactions that have been sent to the given contract at block height 6,837,908, including the deploying transaction.
The reconstructed state holds 759 key/value pairs.
As storing a single variable requires 20,000 gas, storing all values in a single transaction would exceed Ethereum's block size limit of 8,000,000 gas.
The implemented toolbox automatically detects the needed gas and utilizes a proxy contract to split the migration process into multiple transactions.
Finally, the state root of the proxy contract is verifiably identical to the source contract.
Hereby, we could demonstrate the implementation's applicability for token contracts between multiple blockchains.

Every time the proxy contract is addressed after the migration, the call is forwarded to the logic contract.
As a result, the overhead of a delegate contract call is required, resulting in additional gas costs of about 956 gas. 
Therefore, the costs of deploying three smart contracts instead of a single one and requiring additional gas for every method call result in an inferior solution compared the prior migration approach.
However, it is applicable to any contract state size and the additional gas demand is very limited when compared to the costs of storing a single state variable taking 20,000 gas alone~\cite{Wood2014}.

While not intended, the logic contract could be called immediately by any user, potentially changing its state.
As the trusted state maintained on the proxy contract remains unaffected, such actions do not constitute a security threat per se.
Yet, malicious attackers could execute unchecked self destructions if the logic contract exposes such instructions.
As a missing initialization could lead to failures in authorization validation, the deployed contract code needs to be analyzed for such vulnerabilities in advance using tools such as Manticore\footnote{https://github.com/trailofbits/manticore/} or Mythril.
Such added security implications do not apply for single contract migrations, as they remain initialized.

\subsection{Contract Dependencies and Private Chains}
We migrate the smart contract presented in Listing~\ref{lst:simple_contract} to showcase the migration of contracts containing dependencies on other contracts.
The application is first deployed to a  permissioned Ethereum chain using the geth client before it is transferred to a local Quorum node.
Executing the migration process shows that all dependencies are resolved recursively as expected.
Yet, using Quorum as a target chain reveals a limitation regarding verifiability.
The Quroum client supports private smart contracts which share their state only between authorized nodes in the network.
Private transactions are brokered in a dedicated constellation network and stored separately from the public state trie.
While it is possible to migrate contracts to Quorum in a private environment, the current migration implementation cannot validate related state variables.
Though, we could prove smart contract verifiability for public applications using Quorum.

%

%% file: sections/related_work.tex
\section{Related Work}
\label{sec:related_work}
Interactions between multiple systems are referred to as interoperability.
In the context of blockchains, numerous dimensions of interoperability exist.

First, it may refer to the blockchain as an enabler of interoperability between various software systems~\cite{zhang2017}.
Here, the distributed ledger provides a common ground for trusted data exchange and logic execution beyond organizations~\cite{korpela2017}.
As a result, it facilitates process execution for collaborating entities, independent of utilized software implementations~\cite{glaser2017}.
Accessibility and verifiability permit localizing potential implementation issues for liability purposes in distributed systems.

Second, assets such as tokens and cryptocurrencies may be exchanged between different ledgers~\cite{buterin2016interoperability}.
For instance, the Interledger protocol facilitates payment processes and asset exchanges between multiple ledger implementations without requiring any intermediaries or trust in third parties~\cite{Thomas2015}.
By locking funds in escrow contracts, assets are sent via payment channels.
Connector nodes depict interfaces to source and target ledgers.
Following an incentive mechanism that distributes some share of the transferred assets, connectors compete for handling interledger payments.
Transfers are either conducted in an atomic operation including each involved entity and backed by notaries or by utilizing economic incentives, ensuring correct execution.

In contrast, Cosmos creates a network of a hub and multiple zones, where each zone relates to a different ledger~\cite{Kwon2018}.
On the hub, being a distributed ledger itself, consensus is facilitated by the Byzantine-Fault-Tolerant (BFT) Tendermint algorithm that is based on delegated Proof-of-Stake (dPoS)~\cite{Kwon2014}.
Instead of routing assets over multiple hops, they are transferred from one zone via the hub to another zone.
Similar to Cosmos, Polkadot enables interoperability between multiple ledgers in a hub and spoke fashion, but additionally shares security properties of underlying chains~\cite{Wood2016}.
To achieve this, blocks produced in connected chains are validated by respective validator instances.

%% file: sections/discussion.tex
\section{Discussion and Future Work}
While the implemented set of tools for migrating smart contracts and verifying their validity could prove the concept's soundness, potential issues evolving from underlying blockchain implementations remain.
For instance, blockchains utilizing Proof-of-Work (PoW) for finding consensus and leader election suffer from absent finality guarantees~\cite{Buterin2017}.
As a result, the state of a contract could be retrieved from a fork of the blockchain that is not proceeded with due to inferior cumulative PoW.
While validations may succeed as long as they are based on the source fork, the contracts' actual state could differ in such cases.
Hereby, users assume a correct contract migration while in fact the source state has not been recorded in the origin ledger.

The presented concept follows the assumption that source and target ledger share the same address pattern.
Only in this case, key pairs can be reused on the target chain to claim assets, permissions and so forth.
While substituting addresses with those of the target chain could mitigate such issues, it requires prior authentication of all affected users.
In general, this process would be comparable to claiming token ownership on a newly deployed blockchain after participating in an Initial Coin Offering (ICO).
To achieve verifiability, this process should be conducted on the source chain.
Hereby, account ownership can be ensured and information required for validation purposes is available to all participants.

Like any account in Ethereum, smart contracts can hold Ether, its native cryptocurrency.
As an account's balance is directly stored as a value in the account (cf. Section~\ref{subsec:merkle}), it does not affect the contract's storage trie.
Therefore, cryptocurrency held by a contract does not mitigate the validation of transferred state variables.
Yet, linked native currency cannot be migrated to this point.
To transfer such assets, notary networks like Cosmos or Polkadot may be examined for suitability in future implementations (cf. Section~\ref{sec:related_work}).

In Ethereum, smart contracts can emit events for multiple purposes.
For instance, user interfaces may subscribe to events for observing updates in order to display current information.
In other use cases, events are used as a cheap kind of storage, as they require less gas in comparison to persisting information by utilizing state variables~\cite{westerkamp2018}.
While such data is only accessible externally, it may be crucial information depending on the application.
Currently, the presented concept does not consider the migration of events.
As events are bound to the block in which the emitting transaction was mined, their migration to another blockchain constitutes a task for future research.
In order to maintain associated information, both source and target chain are required.
Users should query the source chain till the migration block and the target chain thereafter.
In future implementations, a middleware that forwards requests to respective nodes could ease accessing smart contracts which have been migrated during their life cycle.

%% file: sections/conclusion.tex
\section{Conclusion}
In this paper, we have introduced a concept for facilitating smart contract portability between EVM-compatible blockchains.
Trust requirements in the migrating entity are mitigated by providing verifiability of contract code and state to any participant on the ledger.
Depending on the source contract's size, two distinct concepts have been proposed, enabling migrations of small contracts within a single transaction as well as a set of three contracts to facilitate transferring large contracts in a verifiable fashion.

To showcase the concept's soundness, we presented a toolbox that implements all components of the proposed migration process.
Respective contract runtime code is retrieved directly from the source chain, while all transactions that have been sent to the contract are re-executed locally to reconstruct its current state.
For applications that permit deploying the fetched state in a single transaction, deploycode is generated that stores the contract's logic and initiates its state.
Large contract states are migrated using a dedicated storage contract that is decoupled from its logic in order to maintain verifiability after a migration phase.

We could prove the validity of migrated contracts by utilizing a ERC20 compliant smart contract as well as a sample contract containing references to other contracts deployed on the same chain.
The evaluation of smart contracts using multiple distinct blockchain clients provides evidence of the proposed concept's applicability.